\documentclass[twocolumn,prl,preprintnumbers,amsmath,amssymb,superscriptaddress]{revtex4}

\usepackage{graphicx}
\usepackage{dcolumn}
\usepackage{bm}

\usepackage[colorlinks=true,linktocpage=true,linkcolor=blue,citecolor=blue]{hyperref}

\graphicspath{{figures/}}

\begin{document}

\title{Magnetic Field Induced Polarization Difference between Hyperons and Anti-hyperons}

\author{Yu Guo}
\affiliation{College of Science, China Three Gorges University, Yichang 443002, China.}

\author{Shuzhe Shi}
\affiliation{Department of Physics, McGill University, 3600 University Street, Montreal, QC, H3A 2T8, Canada.} 
\address{Physics Department and Center for Exploration of Energy and Matter, Indiana University, 2401 N Milo B. Sampson Lane, Bloomington, IN 47408, USA.}

\author{Shengqin Feng}
\email{fengsq@ctgu.edu.cn}
\affiliation{College of Science, China Three Gorges University, Yichang 443002, China.}

\author {Jinfeng Liao} \email{liaoji@indiana.edu}
\address{Physics Department and Center for Exploration of Energy and Matter, Indiana University, 2401 N Milo B. Sampson Lane, Bloomington, IN 47408, USA.}

\date{\today}

\begin{abstract}
Recent STAR measurements suggest a difference in the global spin polarization between hyperons and anti-hyperons, especially at relatively low collision beam energy. One possible cause of this difference is the potential presence of in-medium magnetic field. In this study, we  investigate  the  phenomenological viability of this interpretation.  Using the AMPT model framework, we quantify the influence of different magnetic field evolution scenarios   on the size of the polarization difference in a wide span of collision beam energies.  We find that such difference is very sensitive to the lifetime of the magnetic field.  
For the same lifetime, the computed polarization difference only mildly depends on the detailed form of its evolution. Assuming magnetic polarization as the mechanism to enhance anti-hyperon signal while suppress hyperon signal, we phenomenologically extract an upper limit on the needed magnetic field lifetime in order to  account for the experimental data. The so-obtained lifetime values are in a quite plausible ballpark and follow approximately the scaling relation of being inversely proportional to the beam energy. Possible implications on other magnetic field related effects are also discussed. 
\end{abstract}
\maketitle

\section{\label{sec:Intro} Introduction}

Studies of strongly interacting fluid under the influence of rotational motion have attracted significant interests recently, with much excitement particularly triggered by the STAR Collaboration's global polarization measurements in heavy ion collisions~\cite{STAR:2017ckg,Adam:2018ivw}. On quite general ground, one expects  the interplay between macroscopic fluid rotation and microscopic spin of individual particles can lead to many novel effects. For example, individual particle spins will be polarized on average along the global angular momentum. In the context of heavy ion collisions, the colliding system in a non-central collisions carries a large angular momentum along the direction perpendicular to the reaction plane and a global polarization effect shall be expected from the produced hadrons in such collisions~\cite{Liang:2004ph,Gao:2007bc,Voloshin:2004ha,Betz:2007kg,Becattini:2007sr}. More precisely, the angular momentum would turn into interesting vorticity patterns in the QCD fluid and the vorticity structures further induce the particle spin polarization~\cite{thermal_vorticity,Becattini:2014yxa,Becattini:2013vja,Csernai:2013bqa,Csernai:2014ywa,Becattini:2015ska,Becattini:2016gvu,Jiang:2016woz,Shi:2017wpk,Deng:2016gyh,Pang:2016igs,Li:2017slc,Xia:2018tes,Sun:2017xhx,Wei:2018zfb}. The presence of nonzero vorticity can have nontrivial impact on the properties of the underlying matter, such as the phase structures and equation of state~\cite{Yamamoto:2013zwa,Jiang:2016wvv,Ebihara:2016fwa,Chen:2015hfc,Huang:2017pqe,Liu:2017spl,Chernodub:2016kxh,Chernodub:2017ref,Wang:2018sur,Zhang:2018ome,Wang:2018zrn}. If the rotating system consists of chiral fermions, the vorticity can also induce anomalous transport phenomena known as Chiral Vortical Effects~\cite{Son:2009tf,Kharzeev:2010gr,Landsteiner:2011iq,Hou:2012xg,Jiang:2015cva}. For recent reviews, see e.g. 
\cite{Fukushima:2018grm,Kharzeev:2015znc,Liao:2014ava}. 

The global polarization effect measurements by STAR Collaboration in \cite{STAR:2017ckg} show signals for both hyperons and anti-hyperons at the level of a few percent, with a strongly increasing trend toward lower collision energy. The data also clearly demonstrate a visible difference in the polarization between hyperons and anti-hyperons, with $P_{\bar{\Lambda}}>P_\Lambda$ and with such difference also becoming stronger at lower energy. While the average polarization signal could be quantitatively explained by hydrodynamic and transport modelings, the observed difference between hyperons and anti-hyperons remain a puzzle. Efforts were made to investigate various factors that may contribute to such a splitting albeit without conclusive answer~\cite{Becattini:2016gvu,Csernai:2018yok,Muller:2018ibh,Xia:2019fjf,Becattini:2019ntv,Guo:2019mgh}.  At the moment, this is one of the important unresolved challenges within the fluid-vorticity paradigm for the observed global polarization.

One plausible proposal is to take into account an additional polarization (apart from the vorticity-induced effect) due to the existence of in-medium magnetic field which gives opposite polarization effect for hyperons and anti-hyperons~\cite{Becattini:2016gvu,Muller:2018ibh,Guo:2019mgh}. Indeed, there is a very strong initial magnetic field in an off-central heavy ion collision~\cite{Bzdak:2011yy,Deng:2012pc,Bloczynski:2012en,McLerran:2013hla,Gursoy:2014aka,Gursoy:2018yai,Tuchin:2015oka,Inghirami:2016iru,Li:2016tel,Zhong:2014cda,She:2017icp} and if sufficiently long-lived  could provide a considerable amount of magnetic polarization that distinguishes particles from anti-particles. We note in passing that strongly interacting matter under strong magnetic field has in itself been a very active topic of significant interests with many developments (see recent reviews in e.g. \cite{Kharzeev:2015znc,Fukushima:2018grm,Liao:2014ava,Miransky:2015ava,Hattori:2016emy}).

The main objective of the present study is to  explore the phenomenological viability of such a magnetic-field-based interpretation for the observed difference in hyperon/anti-hyperon global polarization.  Using the AMPT model framework and incorporating both rotational and magnetic polarization effects, we will quantify the influence of different magnetic field evolution scenarios   on the polarization difference in a wide span of collision beam energies. We will use the polarization difference to phenomenologically extract an upper limit on the needed magnetic field lifetime in order to fully account for the experimental data. We will discuss the behavior of  so-obtained lifetime values and discuss possible implications on other magnetic field related effects.

\section{\label{sec:Method} Formalism}

In this part we provide a detailed description of the formalism we use to compute the $\Lambda$ and $\bar\Lambda$ polarization. For the overall bulk matter created in the collisions, we use the transport model AMPT~\cite{Lin:2004en,Lin:2014tya} for a number of reasons. First, it provides a reasonable description of the bulk collective dynamics such as soft particles' yields, transverse momentum spectra and flow observables.  We use the same setup  as in \cite{Lin:2014tya} which demonstrated very good agreement with experimental data for Au+Au collisions at RHIC. Furthermore it can be used for a wide span of collision beam energies. Another advantage   is that it allows explicit tracking of every parton or hadron's motion during the evolution and of each final state hadron's formation. This allows a relatively straightforward procedure to extract the system's vorticity structure as well as to incorporate the spin polarization effect upon the hadron formation. The AMPT model  was first extended to compute vorticity structures in  \cite{Jiang:2016woz} and  later widely used for polarization studies~\cite{Shi:2017wpk,Li:2017slc,Xia:2018tes}. From AMPT simulations one obtains the four velocity distribution $u^\mu(x)$ as well as energy density distribution $\epsilon(x)$ in space-time $x=(t,\vec x)$ across the system, which can be further used to evaluate various quantities of interest.    

The rotational polarization effect on particle spin in a relativistic fluid can be determined from the   thermal vorticity $\varpi_{\mu\nu}$ defined as~\cite{thermal_vorticity,Becattini:2014yxa}:  
\begin{eqnarray} \label{eq_varpi}
 \varpi_{\mu \nu} = - \frac{1}{2}(\partial_\mu \beta_\nu - \partial_\nu \beta_\mu) 
\end{eqnarray} 
 where $\beta_\mu = u_\mu/T$ with $T = 1/\beta$ the local temperature. A  related quantity is the kinetic vorticity defined by 
$ \Omega_{\mu\nu}= - \frac{1}{2}(\partial_\mu u_\nu - \partial_\nu u_\mu) $.
 Obviously 
 $\varpi_{\mu \nu} = \beta \left\{  \Omega_{\mu\nu} -  \left[  (\beta \partial_\mu T) u_\nu -     (\beta \partial_\nu T) u_\mu  \right]  \right\}$.  The thermal vorticity differs from the $\Omega_{\mu\nu}/T$   by terms containing gradients of temperature,  
$\sim (\beta \partial_\mu T)= [(\partial_\mu T)/T]$. While straightforward to evaluate in hydrodynamic models, such terms are trickier to compute in transport models. As a proxy, we use the energy density $\epsilon$ to evaluate such terms via $(\partial_\mu T)/T=(\partial_\mu \epsilon)/(4\epsilon)$ with underlying assumption $\epsilon \propto T^4$. Such gradient terms make non-negligible contributions and should be taken into account.

 We now discuss the calculation of particle polarization e.g. for the hyperons and anti-hyperons. In the case that polarization {\em solely} comes from vorticity, one has the following  ensemble-averaged spin 4-vector of the produced $\Lambda$ and $\bar{\Lambda}$   determined from the local thermal vorticity at its formation location, as~\cite{thermal_vorticity,Becattini:2014yxa,STAR:2017ckg,Becattini:2016gvu,Li:2017slc}: 
\begin{eqnarray} \label{eq_neutral}
S^\mu =  - \frac{1}{8m}\epsilon^{\mu\nu\rho\sigma}p_\nu \varpi_{\rho\sigma}  
\end{eqnarray} 
where $p^\nu$ is the four-momentum and $m$ the   mass of the produced hyperons/anti-hyperons. Past calculations solely based on the vorticity-induced polarization can not describe the observed difference between signals of $\Lambda$ and $\bar{\Lambda}$. In fact, as we will show later, the polarization effect from just the vorticity of fluid rotation would be larger for $\Lambda$ than $\bar{\Lambda}$, quite the opposite to data, due to a subtle effect related to particle formation timing. 

The existence of a magnetic field could indeed induce a difference in the spin polarization between $\Lambda$ and $\bar{\Lambda}$ due to their opposite magnetic moments. Under the presence of electromagnetic fields $F_{\mu\nu}$, the spin 4-vector formula will become different from that in Eq.(\ref{eq_neutral}) and should be given instead by the following~\cite{Becattini:2016gvu}: 
\begin{eqnarray} \label{eq_P_B}
\tilde{S}^\mu =  - \frac{1}{8m}\epsilon^{\mu\nu\rho\sigma}p_\nu 
\left[  \varpi_{\rho\sigma}  \mp 2 (eF_{\rho\sigma})  \mu_\Lambda/T_f  \right ]
\end{eqnarray} 
where $\mu_\Lambda = \frac{0.613}{2m_N}$ is the absolute value of the hyperon/anti-hyperon magnetic moment, with $m_N=938\rm MeV$ being the nucleon mass.  $T_f$ is the local temperature upon the particle's formation.  In the case with nonzero electromagnetic field, there will be a difference between  $\Lambda$ and $\bar{\Lambda}$  spin polarization due to the second term in the above. Here we focus on the electromagnetic field component that is most relevant to the global polarization effect, namely $B_y = F_{31} = - F_{13}$ along the out-of-plane direction which is also the direction of global angular momentum.  It should be noted that the above Eq.(\ref{eq_P_B}) assumes local equilibrium of polarization under electromagnetic fields. In the rather dynamical environment of heavy ion collisions, particle polarization may not necessarily relax instantaneously to the expected value and off-equilibrium corrections could be important. This is an interesting problem for future study. One important caveat for comparison with experimental data is the influence of secondary decays on the measured hyperon polarization. Two important recent studies~\cite{Xia:2019fjf,Becattini:2019ntv} have excluded a major role of such decay contributions and therefore justified the application of Eq.(\ref{eq_P_B}) for primary hadrons in our study as a very good approximation.       

 In (non-central) heavy ion collisions, there is a strong initial magnetic field $eB_0$ arising from the fast-moving spectator protons, which has been very well studied~\cite{Bzdak:2011yy,Deng:2012pc,Bloczynski:2012en}. The key issue here is whether such a magnetic field would survive long enough to have nonzero impact around the freeze-out time for hadron formation.  There are proposals for certain mechanisms that could provided relatively long-lived late time magnetic field, e.g. by way of medium induction~\cite{McLerran:2013hla,Gursoy:2014aka,Gursoy:2018yai,Tuchin:2015oka,Inghirami:2016iru,Li:2016tel} or by rotating fluid with nonzero charge density~\cite{Guo:2019mgh}.   Nevertheless currently the magnetic field time evolution in heavy ion collisions is rather uncertain~\cite{Huang:2017tsq}. Alternatively, one may turn this around (as suggested in \cite{Muller:2018ibh}) and consider the splitting between $\Lambda$/$\bar{\Lambda}$ polarization  as a way to put an empirical constraint on the size of potentially existing late time magnetic field. In the present study, we further exploit this line of thought and address the following question: what kind of magnetic field time evolution $B_y(\tau)$ would be needed, if the observed polarization difference  would be entirely attributed to the in-medium magnetic field?    
 
\begin{figure}[htb!]
\includegraphics[width=7cm]{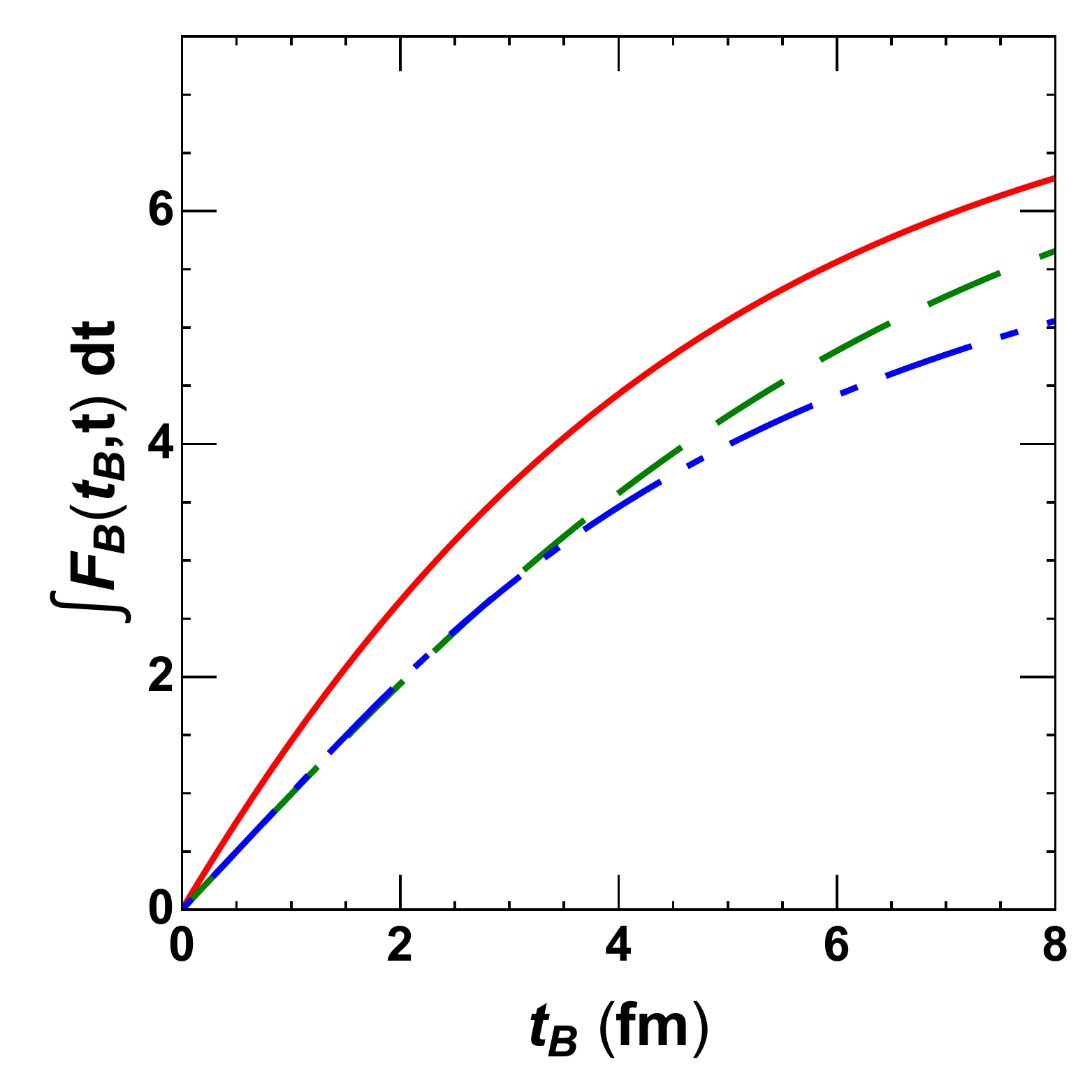}
\caption{\label{fig0} (color online) The time integrated value $\int F(t_B,t) dt $ (for $t$ from $0\sim 8$ fm/c time) as a function of $t_B$ for different magnetic field time evolution: type-1 (red solid curve), type-2 (green dashed curve) and type-3 (blue dash-dotted curve). See text for details.}
\end{figure}
  
 In order to study this question, and given the uncertainty about the time dependence, we phenomenologically investigate this problem by assuming $B_y(t;\vec{x})=B_0(\vec{x})\cdot F_B(t_B,t)$ and studying three different kinds of parameterization for $F_B$ that have been adopted in the literature for various studies of magnetic field effects:  \\ 
Type-1: , $F_B(t_B,t)\equiv\frac{1}{1+(t-t_0)^2/t_B^2}$ (see e.g.~\cite{Jiang:2016wve,Shi:2017cpu}); \\
Type-2: $F_B(t_B,t)\equiv\frac{1}{\left[ 1+(t-t_0)^2/t_B^2 \right]^{3/2}}$ (see e.g.~\cite{Deng:2012pc}); \\ 
Type-3:  $F_B(t_B,t)\equiv  e^{-|t-t_0|/t_B} $(see e.g.~\cite{Muller:2018ibh}). \\
 In all these parameterizations, the  $t_B$ is the essential magnetic field lifetime parameter that controls how rapidly the magnetic field would decrease with time. Note however due to their different functional forms, the same $t_B$ value gives slightly different magnetic field evolution. To give an idea of such difference, we show in Fig.~\ref{fig0} the time integrated value $\int F(t_B,t) dt $ (for $t$ from $0\sim 8$ fm/c time) as a function of $t_B$ for comparing these three types of evolution.   Defining $t=0$ as the time point of the very initial contact of the collision process,  the $t_0\equiv R_{A}/(\gamma_{beam} v_{beam})$ is the time for full overlap of the two colliding nuclei, with $R_{A}$ being the nuclear radius, $v_{beam}$ and $\gamma_{beam}$   being the beam speed and the corresponding Lorentz factor.  Note this is important particularly for collisions at low beam energy. The initial magnetic field value $B_0(\vec{x})$ is determined from event-by-event calculations with Monte-Carlo Glauber simulations as in e.g. \cite{Bloczynski:2012en}. Note this field strongly depends on beam energy, following a trend $B_0\propto \sqrt{s_{NN}}$. For example, at the center point $\vec{x}=0$, the initial  strength $\frac{eB_0(\vec{x}=0)}{m_\pi^2}$ (where $m_\pi$ is the pion mass) equals $0.222,\ 0.282,\ 0.383,\  0.528,\ 0.764,\ 1.235,\ 3.922$ for beam energy $\sqrt{s_{NN}}=11.5,\ 14.5,\ 19.6,\ 27,\ 39,\ 63,\ 200\ \rm GeV$ respectively. These values are determined from simulating proton distributions in the initial conditions and are consistent with other calculations. In this study we focus on the $(20\sim50)\% $ centrality class of AuAu collisions which correspond to the STAR measurements in \cite{STAR:2017ckg} and we simulate $10^6$ or more AMPT events for each given beam energy to ensure enough statistics. The hyperon and anti-hyperon polarization results are then computed from Eq.(\ref{eq_P_B}) for each type of magnetic field time evolution with a chosen lifetime parameter. We present the detailed results from such study in the next section.

\section{\label{sec:Results}  Results}

\begin{figure}[htb!]
\includegraphics[width=8cm]{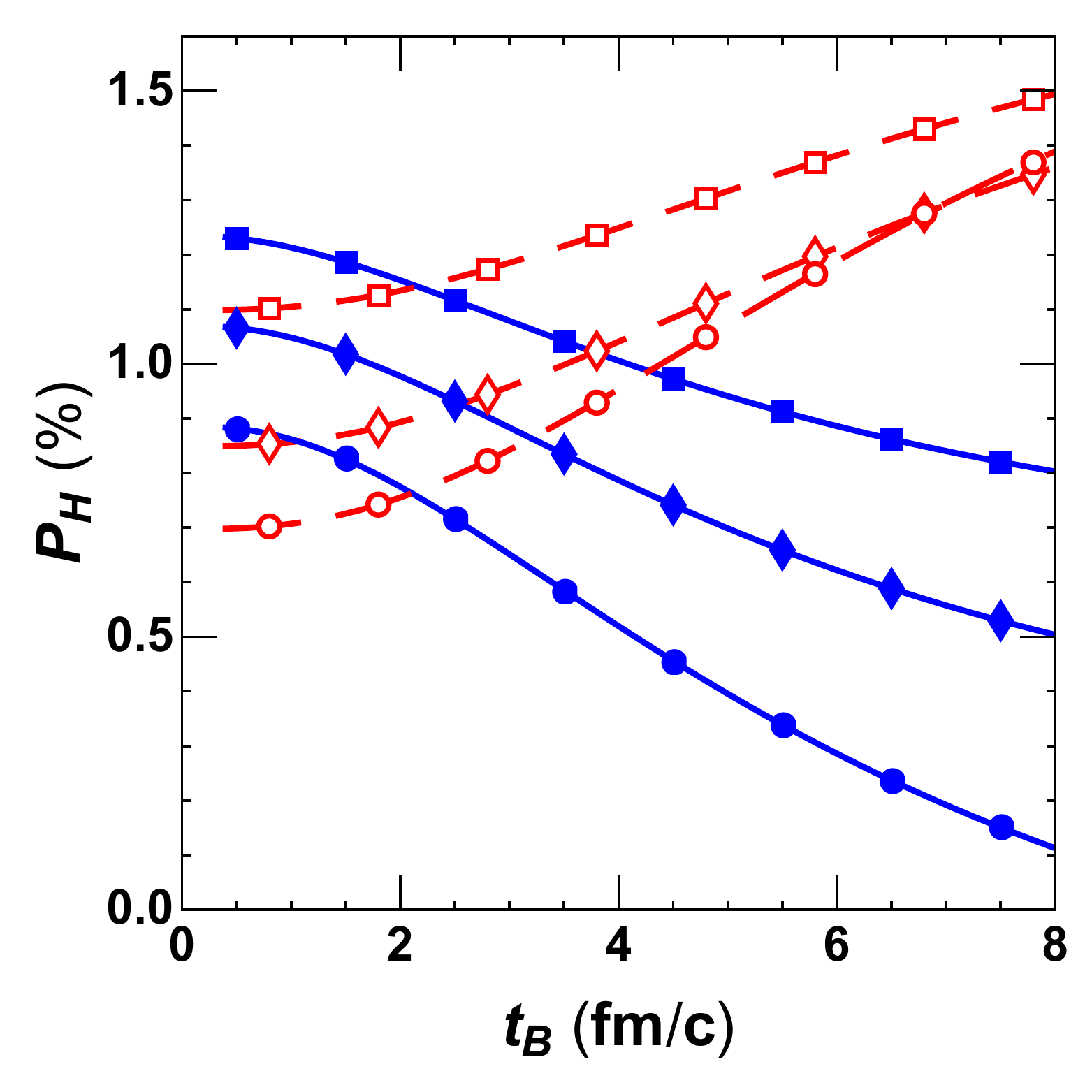}
\caption{\label{fig1} (color online) The dependence on magnetic field lifetime parameter $t_B$ of the global polarization signals $P_H$ for hyperons ( $H \to \Lambda$, blue solid curves with filled symbols) and anti-hyperons ( $H \to \bar{\Lambda}$, red dashed curves with open symbols)  at  beam energy 
$\sqrt{s_{NN}}=$19.6 (square), 27 (diamond), 39 (circle) GeV respectively.  }
\end{figure}

\begin{figure}[htb!]
\includegraphics[width=8cm]{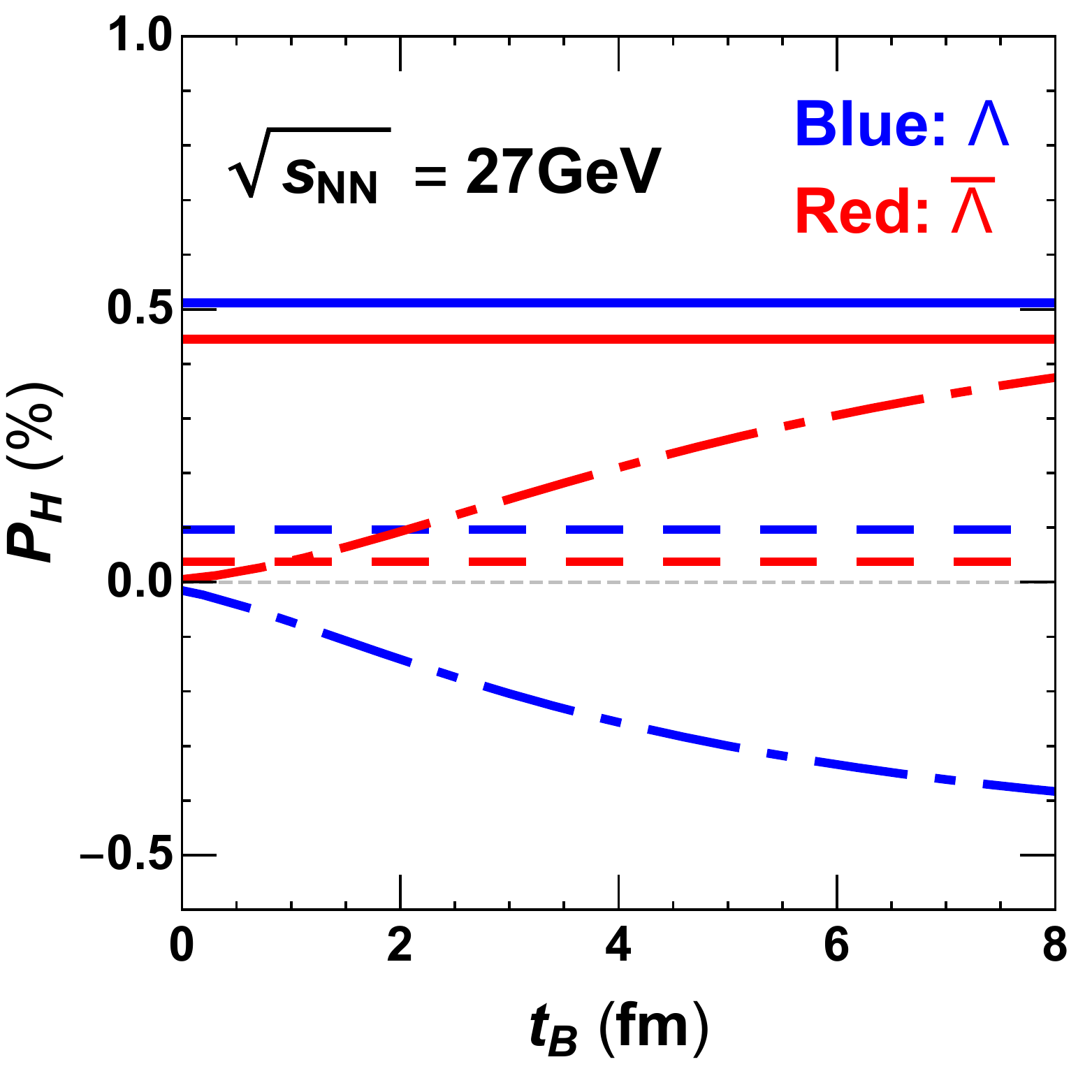}
\caption{\label{fig1b} (color online)  Different contributions to the $\Lambda$ and $\bar{\Lambda}$ polarization from kinetic vorticity term (solid curves), from temperature gradient term (dashed curves) and from magnetic field term (dash-dotted curves) respectively. See text for details.}
\end{figure}

As a first step, let us examine how the key parameter, magnetic field lifetime $t_B$, would influence the polarization observable. To do this, we vary this parameter (for each given type of time evolution) and examine how the obtained  global polarization signals of $\Lambda$ and $\bar{\Lambda}$ would change. In Fig.~\ref{fig1} we show such results for collisions at beam energy $\sqrt{s_{NN}}=19.6,\ 27,\ 39\ \rm GeV$ respectively. 
In Fig.~\ref{fig1b}, we also show and compare the different contributions to the $\Lambda$ and $\bar{\Lambda}$ polarization from kinetic vorticity term, from temperature gradient term and from magnetic field term, suggesting a dominant role of kinetic vorticity and a non-negligible temperature gradient contribution.   
As one can see, with increasing magnetic field lifetime (which means stronger magnetic field at late time in the collisions), the $P_{\bar{\Lambda}}$ steadily increases while the $P_\Lambda$ decreases at all collision energies. With long enough $t_B$, eventually the $P_{\bar{\Lambda}}$ always becomes larger than $P_\Lambda$. The occurrence of ``crosspoint'' (where $P_{\bar{\Lambda}}=P_\Lambda$) requires longer lifetime at lower beam energy. Another interesting observation  is that when $t_B\to 0$ (meaning no magnetic field and only vorticity-induced effect), the hyperons actually have a larger polarization than the anti-hyperons. The origin of this difference is due to an interplay between formation timing and vorticity evolution~\cite{Shi:2017wpk}. We've explicitly checked in AMPT simulations that the averaged production time of hyperons is indeed earlier than that of anti-hyperons and thus the hyperons ``pick up'' a stronger vorticity-induced polarization effect upon formation because of a larger vorticity value  at earlier time.  Note we have not considered a possible finite relaxation time for the particle polarization in the magnetic field, which may reduce the magnitude of the  suppression/enhancement on  $\Lambda$/$\bar{\Lambda}$ polarization obtained in the present study.

\begin{figure}[htb!]
\includegraphics[width=8cm]{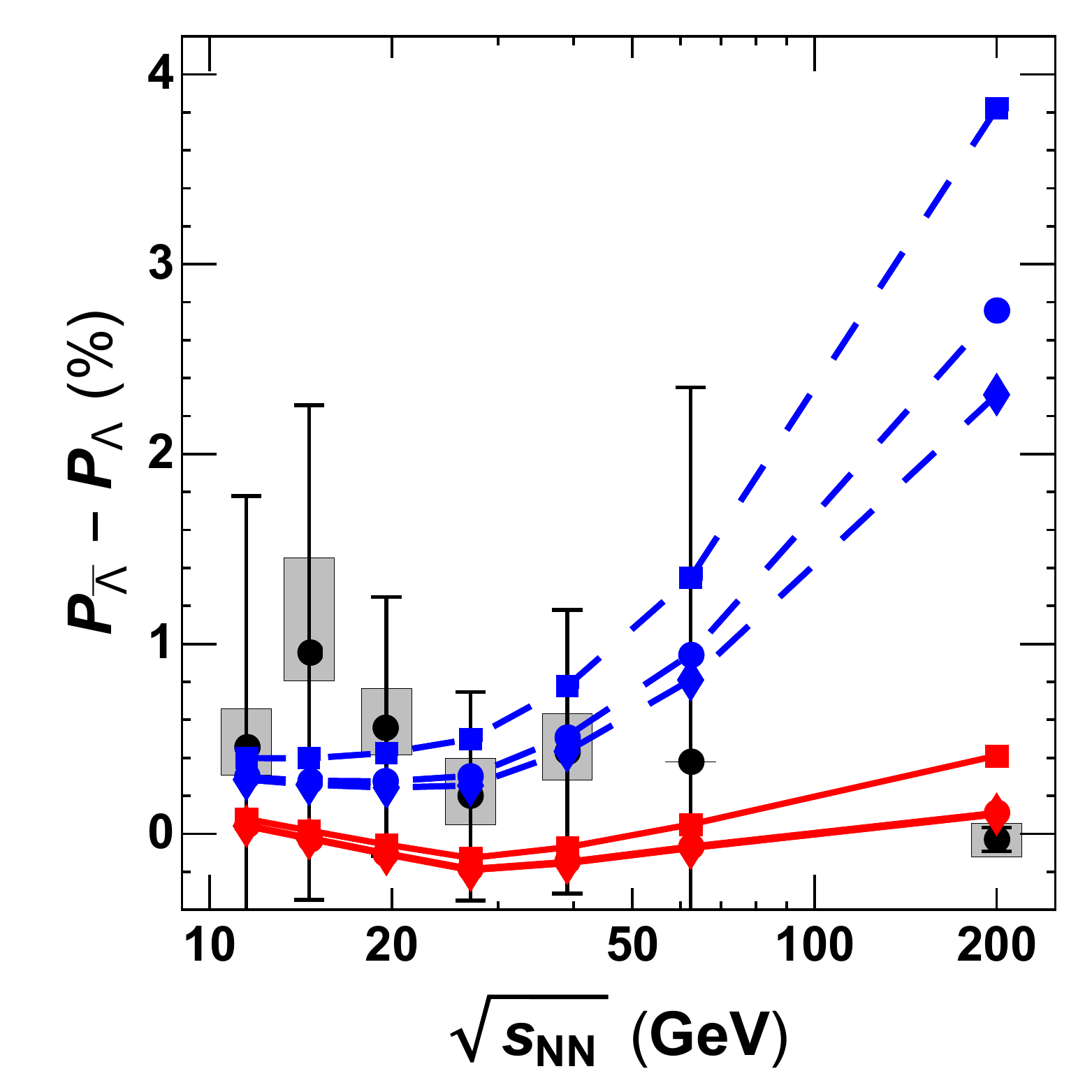}
\caption{\label{fig2} (color online) The difference $\Delta P= P_{\bar{\Lambda}}-P_\Lambda$  versus collision beam energy, for the type-1 (square), type-2 (diamond) and type-3 (circle) time dependence. The red solid curves are for $t_B=1\rm fm$ while the blue dashed curves are for $t_B=4\rm fm$. The black circles with error bars are STAR experimental data from \cite{STAR:2017ckg,Adam:2018ivw}. }
\end{figure}

\begin{figure*}
\includegraphics[width=5cm]{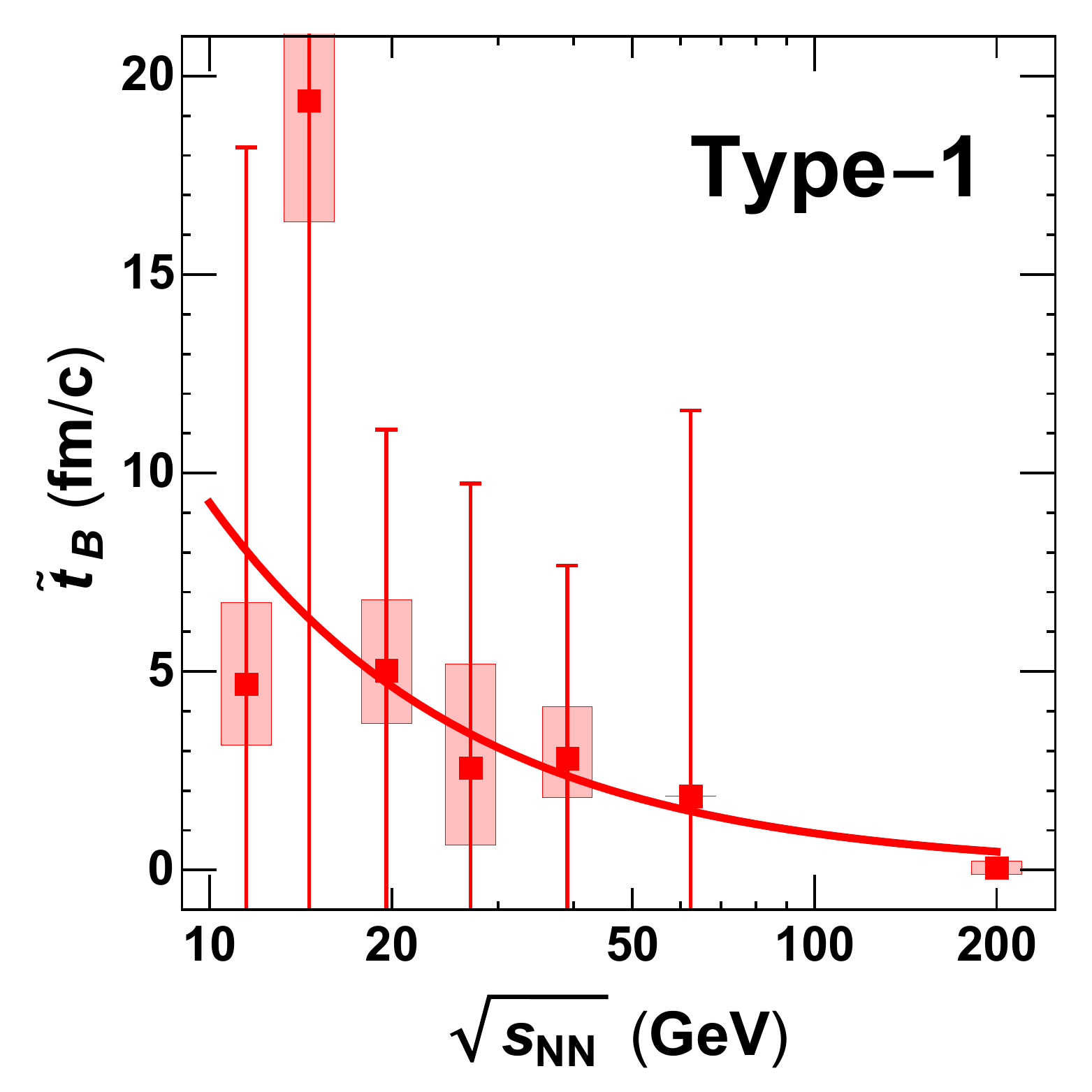} \hspace{0.5cm}
\includegraphics[width=5cm]{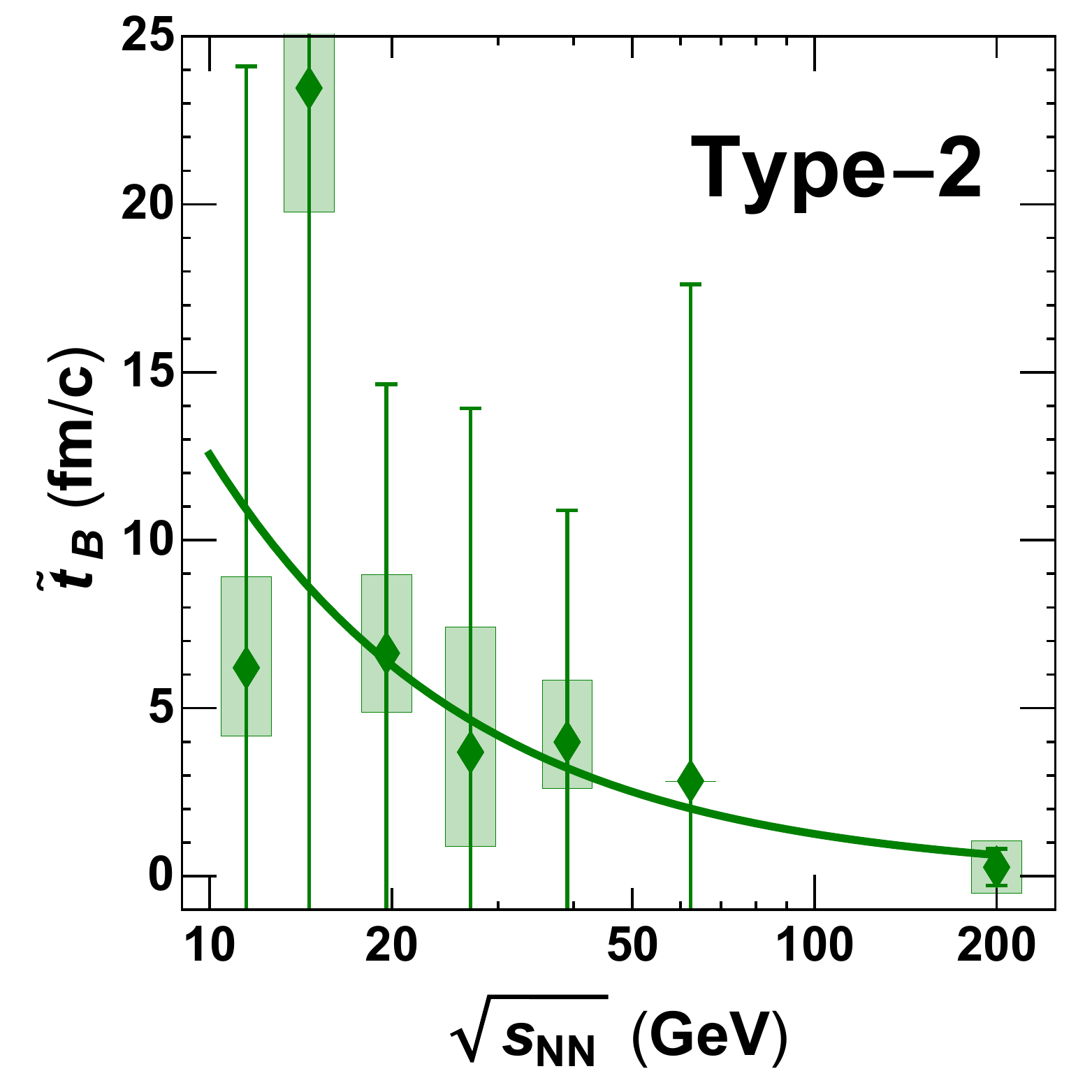} \hspace{0.5cm}
\includegraphics[width=5cm]{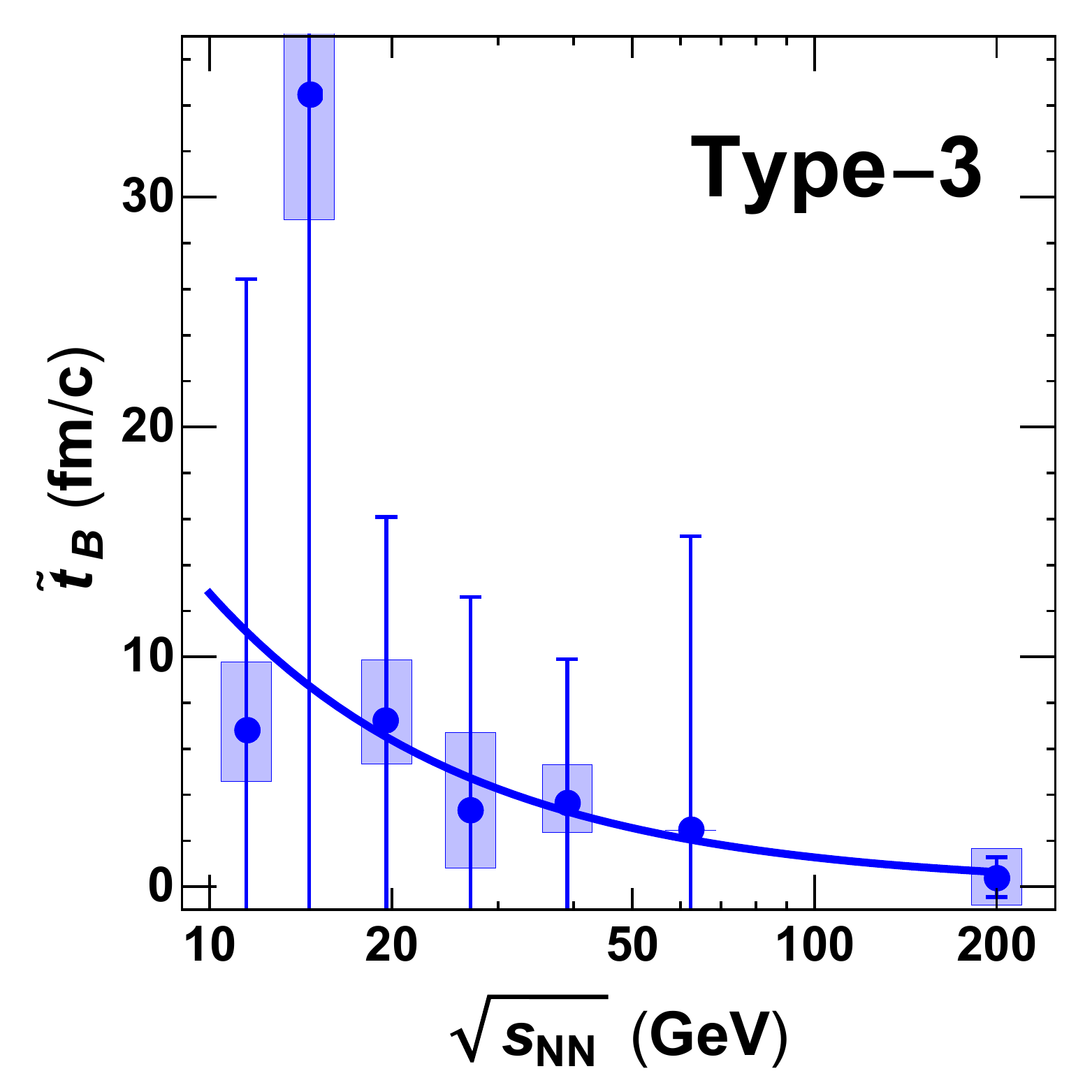}
\caption{\label{fig3} (color online) The optimal value of magnetic field lifetime parameter $\tilde{t}_B$ extracted from polarization splitting $\Delta P$ data for a range of collision beam energy $\sqrt{s_{NN}}$. The left, middle and right panels correspond to the type-1, 2 and 3 forms of magnetic field time evolution. The solid curves are from fitting analysis with a formula $\tilde{t}_B = \frac{A}{\sqrt{s_{NN}}}$. The error bars are converted from the corresponding errors of experimental data in \cite{STAR:2017ckg,Adam:2018ivw}. }
\end{figure*}

It is interesting to check the sensitivity of such magnetic field induced splitting $\Delta P= (P_{\bar{\Lambda}}-P_\Lambda)$   to the details of the time evolution. To do that, we evaluate and compare the $\Delta P$ values computed from the three types of time dependence, with the results shown in Fig.~\ref{fig2}. There, we plot $\Delta P$  versus beam energy $\sqrt{s_{NN}}$ for the type-1,2,3 magnetic fields with two choices of lifetime $t_B$ (a shorter one of 1 fm/c and a longer one of 4 fm/c ). The comparison demonstrates that the magnetic field induced splitting $\Delta P= (P_{\bar{\Lambda}}-P_\Lambda)$, while most sensitive to the parameter $t_B$, also mildly depends on the detailed form of the time evolution.  It is also clear that for the same $t_B$ value, the magnetic field effect is stronger at higher beam energy, simply due to its larger peak value $B_0$. We also show the STAR measured splitting on the same plot, which indicates that a longer lifetime is required for describing the $\Delta P$ at lower beam energy.

\begin{figure}[htb!]
\includegraphics[width=7cm]{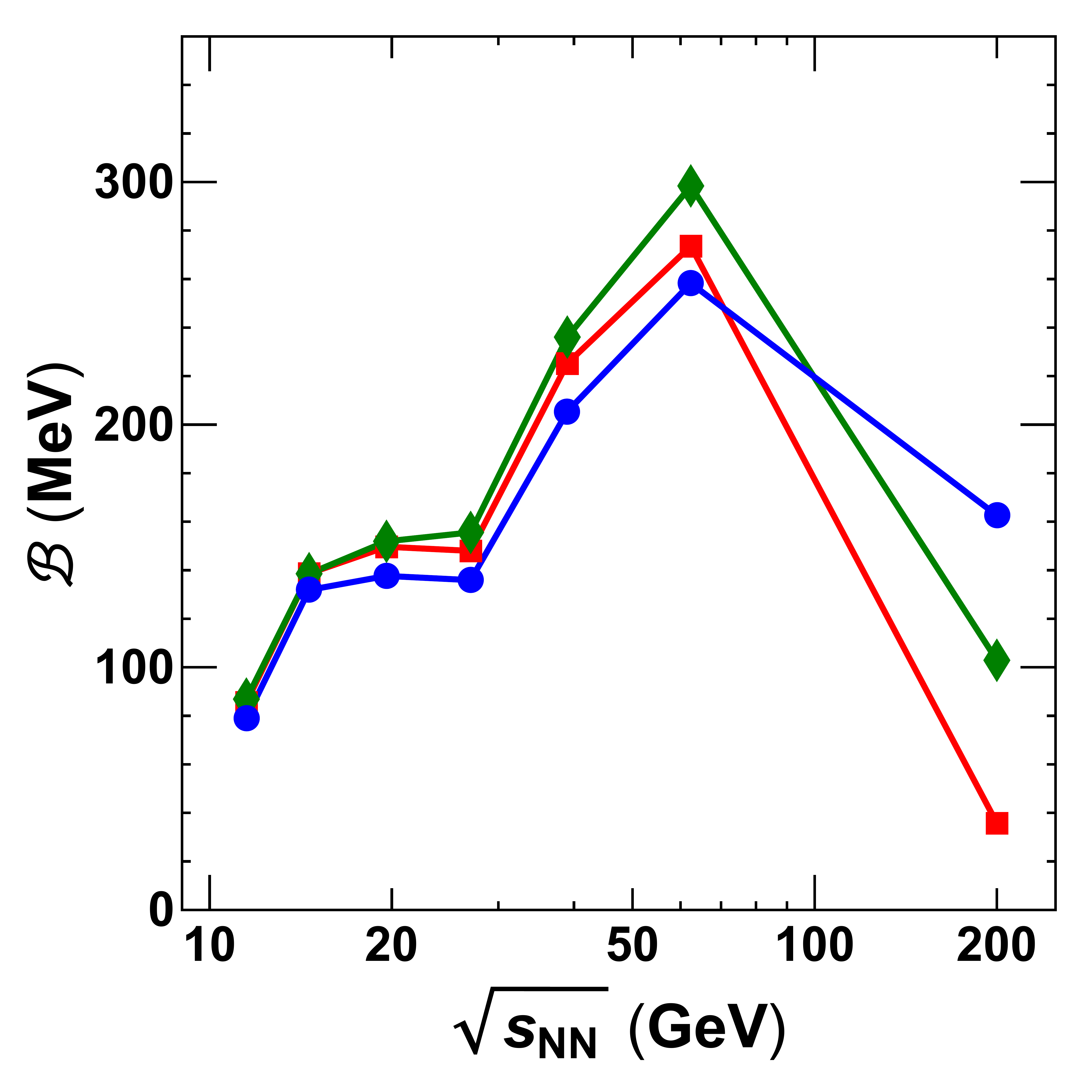}
\caption{\label{fig4} (color online)   The time-integrated magnetic field strength ${\cal B}\equiv \int (eB_y)dt$ at the center point $\vec{x}=0$ as a function of beam energy, for  the type-1 (red square), type-2 (green diamond) and type-3 (blue circle) time dependence with optimized parameter $\tilde{t}_B$ from polarization splitting. }
\end{figure}

We now use the experimental data for $\Delta P$ as a way to constrain the magnetic field lifetime parameter. At each beam energy, we find the optimal value of $\tilde{t}_B$ that would give the amount of measured splitting. This allows us to extract from data the preferred   lifetime as a function of beam energy in a scenario that the splitting is caused by such magnetic field. The results for each of the type-1, 2, 3 (as left, middle, right panels) are shown in Fig.~\ref{fig3}. The error bars are converted from the corresponding experimental data error bars, which at the moment are substantial but may be significantly reduced in upcoming RHIC Beam Energy Scan II measurements~\cite{Bzdak:2019pkr}.  Common to all three types, the needed lifetime $\tilde{t}_B$  decreases  with beam energy $\sqrt{s_{NN}}$. For example, $\tilde{t}_B \sim 5 \rm fm/c$ for $\sqrt{s_{NN}}=11.5\rm GeV$ and $\tilde{t}_B \sim 0.5 \rm fm/c$ for $\sqrt{s_{NN}}=200\rm GeV$.  We note that these numbers may be quite plausible. To quantify such dependence, we perform a fitting analysis, with the dependence $\tilde{t}_B = \frac{A}{\sqrt{s_{NN}}}$. Such a scaling formula is based on Lorentz contracted time for the passing-through period between two nuclei, $\tilde{t}_B \propto \frac{R_A}{\gamma} \propto \frac{1}{\sqrt{s_{NN}}}$. The fitting curves are shown in Fig.~\ref{fig3} as solid curves, with the $\chi^2$-optimized parameter $A=92$ for type-1, $A=125$ for type-2 and $A=128$ for type-3 (all bearing unit of $
\rm GeV\cdot fm/c$).  An averages over these three types of time dependence in a  (perhaps naive) statistical way would suggest $\tilde{t}_B = \frac{A}{\sqrt{s_{NN}}}$ with $A=115\pm 16\ \rm GeV\cdot fm/c$.  Interestingly, this is considerably longer than the expected vacuum magnetic field lifetime without any medium effect, which could be estimated by $t_{vac}\simeq \frac{2R_A}{\gamma}\simeq \frac{26\ \rm GeV\cdot fm/c}{\sqrt{s_{NN}}}$. Such extended magnetic field lifetime, as indicated by polarization difference, may imply a considerable role of the medium feedback on dynamical magnetic field evolution.

A magnetic field, in addition to inducing splitting between $\Lambda$/$\bar{\Lambda}$ polarization, can also lead to various other interesting phenomena~\cite{Kharzeev:2015znc,Fukushima:2018grm}. Many of these effects are dependent on the time-accumulative effect of the magnetic field. With the above analysis of the magnetic field time evolution  based on polarization splitting, we compute a related quantity, the time-integral of magnetic field strength ${\cal B}\equiv \int (eB_y)dt$ at the center point $\vec{x}=0$. This is computed at each beam energy and for each type of time evolution (along with optimized parameter $\tilde{t}_B$), with the results shown in Fig.~\ref{fig4}.  We note that this provides an estimate of the upper limit for accumulative magnetic field strength based on polarization splitting data, which would be useful for constraining other effects arising from the magnetic field. These results suggest that the time-integrated in-medium magnetic field could be considerable and much exceed  the time-integrated vacuum magnetic field as estimated in e.g.~\cite{Asakawa:2010bu}. As shown by Anomalous-Viscous Fluid Dynamics (AVFD) simulations~\cite{Jiang:2016wve,Shi:2017cpu}, an in-medium magnetic field of this scale could make a substantial contribution to the signal of Chiral Magnetic Effect (CME).  It is also interesting to note that the potential CME signal as extracted by STAR Collaboration~\cite{Adamczyk:2014mzf} via the so-called H-correlator from two-component decomposition analysis~\cite{Bzdak:2009fc,Bzdak:2012ia} shows a very similar trend in its beam energy dependence, first increasing and then decreasing with a peak around $\sqrt{s_{NN}}=(40\sim60) \ \rm GeV$ region.

\section{\label{sec:Summary}  Summary}

In summary, we have quantitatively investigated the magnetic field as a probable cause of the observed difference in global polarization between hyperons and anti-hyperons. Using the AMPT model framework, we have quantified the influence of different magnetic field lifetime and time dependence on the size of the splitting in a wide span of collision beam energies. 

Our main findings include: (1) At all beam energies, a longer the magnetic field lifetime leads to a larger polarization for anti-hyperons while a smaller polarization for hyperons; (2) the lifetime parameter sensitively controls the size of the splitting, which is also mildly dependent on the precise form of magnetic field evolution;  (3) The needed magnetic field lifetime in order to fully account for the observed splitting $\Delta P$ is in a plausible ballpark and strongly decreases from low to high beam energy, ranging from a few fm/c  at the low end of RHIC BES energy to a fraction of one fm/c at top RHIC energy; (4) The so-extracted magnetic field lifetime follows approximately the scaling relation of being inversely proportional to the beam energy.   

To conclude, the interpretation of observed polarization difference in terms of lasting magnetic field could be a plausible one and the required magnetic field lifetime appears not impossible. In the present study, we've not addressed the question of precisely how the magnetic field should evolve, which would require  dynamical simulations that solve the magnetic field evolution from Maxwell equations. This would be an important and interesting next step to examine the viability of such interpretation, which we shall carry out in a future study.    
 \vspace{0.05in}

This work is partly supported by NSFC Grant No. 11875178 and No. 11735007, by NSF Grant No. PHY-1913729, and by the U.S. Department of Energy, Office of Science, Office of Nuclear Physics, within the framework of the Beam Energy Scan Theory (BEST) Topical Collaboration. SS is grateful to the Natural Sciences and Engineering Research Council of Canada for support. JL thanks the Institute for Advanced Study of Indiana University for partial support. 
\vspace{-0.2in}


\end{document}